\documentclass[reprint,amsmath,amssymb, aps, prd]{revtex4-1} 
\usepackage{graphicx}
\usepackage{dcolumn}
\newcommand{\correction}[1]{#1}
\begin{document} 

          \title{Elastic $pp$-scattering at $\sqrt s$=7 TeV 
               with the genuine Orear regime and the dip} 

\author{I.M. Dremin}
 \email{dremin@lpi.ru}
\author{V.A. Nechitailo}%
 \altaffiliation[Also at ]{Institute of Theoretical and Experimental Physics, Moscow, Russia}

\affiliation{%
 P.N.~Lebedev Physical Institute, Moscow 119991, Russia
}%

\hfill                        In memory of Igor Andreev                     

\begin{abstract}
The unitarity condition unambigously requires the Orear region to appear in
between the diffraction cone at low transferred momenta and hard parton 
scattering regime at high transferred momenta in hadron elastic scattering.
It originates from rescattering of the diffraction cone processes.
It is shown that such region has been observed  in the differential cross 
section of the elastic $pp$-scattering at $\sqrt s$=7 TeV. The Orear region is 
described by exponential decrease with the scattering angle and imposed on it 
damped oscillations. They explain the steepening at the end of the diffraction 
cone as well as the dip and the subsequent maximum observed in TOTEM data. The 
failure of several models to describe the data in this region can be understood 
as improper account of the unitarity condition. It is shown that the real part
of the amplitude can be as large as the imaginary part in this region.
The overlap function is calculated and shown to be small outside the 
diffraction peak. Its negative sign there indicates the important role of phases
in the amplitudes of inelastic processes. 
\end{abstract}


\maketitle

\section{Introduction}

The TOTEM collaboration has published \cite{totem,*totem1} experimental results on the
differential cross section of the elastic $pp$-scattering at the total cms 
energy $\sqrt s$=7 TeV. Among the
most interesting features they observe the steepening of the diffraction cone
near the squared transferred momentum 0.3 GeV$^2$, the dip at 0.53 GeV$^2$ and
the maximum at 0.7 GeV$^2$. We explain them as resulting from the rigorous
requirements of the unitarity condition. It prescribes the Orear regime 
characterized by exponential decrease with the scattering angle to start at 
transferred momenta just above the diffraction cone. The damped oscillations 
imposed on it lead to the dip in the differential cross section. No particular 
model has been used.

At the same time there exist several models mostly based on reggeon approach.  
Their predictions are extensively cited in \cite{totem,*totem1}. 
\correction{Being rather successful in the diffraction cone,} they fail to 
describe the new data quantitatively beyond the diffraction peak. This 
demonstrates that the unitarity condition is not properly accounted  there
in these models. Since then some other models were proposed \cite{ugal, sel}. 

At the end of 60s the very first experimental data on elastic $pp$- and $\pi p$-
scattering were obtained at energies between 6.8 and 19.2 GeV in the laboratory 
system \cite{cocc, orear, alla,*alla1}. They showed that just after the diffraction 
cone which behaved as a Gaussian in the scattering angle there was observed 
the exponentially decreasing with the angle behavior which
was called as the Orear regime in the name of its investigator \cite{orear}.
Some indications on the shoulder appearing at the beginning of this region 
(evolved later to the dip at the higher ISR energies) were also obtained.
The special session was devoted to these findings at the 1968 Rochester conference in 
Wien. 

The theoretical indications on the possibility of such regime were obtained 
even earlier \cite{hove, amati, cott} but the results did not fit new
experimental findings.

At the same time the simple theoretical explanation based on rigorous 
\correction{ model-independent}
consequences of the unitarity condition was proposed \cite{anddre, anddre1} 
and a careful fit to experimental data showed good quantitative agreement 
with experiment \cite{adg}.

We follow these ideas to demonstrate that they are also applicable to the recent 
data of the TOTEM collaboration at the LHC at energies as high as 7 TeV.

\section{Theoretical description}

The elastic scattering proceeds mostly at small angles. The diffraction peak
has a Gaussian shape in the scattering angles or exponentially decreasing as
the function of the transferred momentum squared  
\begin{equation}\
\frac {d\sigma }{dt}/\left( \frac {d\sigma }{dt}\right )_{t=0}=e^{Bt}\approx 
e^{-Bp^2\theta ^2},
\label{diff}
\end{equation}
where the four-momentum transfer squared is
\begin{equation}
t=-2p^2(1-\cos \theta )\approx -p^2\theta ^2 \quad 
(\theta \leq \theta _d \ll 1)
\label{trans}
\end{equation}
with $p$ and $\theta $ denoting the momentum and the scattering angle in the 
center of mass system and $B$ known as the diffraction slope. 

At large energies the forward scattering amplitude has a small real part as
known from the dispersion relations \cite{drna, blha}. Therefore,
in the first approximation, it is reasonable to assume that its real part is 
negligible within the diffraction peak $\theta \leq \theta _d$. Then the 
elastic scattering in this region can be described by the amplitude 
\begin{equation}
A(p,\theta )\approx 4ip^2\sigma _te^{-Bp^2\theta ^2/2}
\label{ampl}
\end{equation}
with a proper optical theorem normalization to the total cross section 
$\sigma _t$ in the forward direction. \correction{ We stress that  Eq. (\ref{ampl})
follows directly from experimental results and does not appeal to
any particular model}.

Let us have a look at the unitarity condition which is
\begin{widetext}
\begin{equation}
{\rm Im}A(p,\theta )= I_2(p,\theta )+F(p,\theta )= 
\frac {1}{32\pi ^2}\int \int d\theta _1
d\theta _2\frac {\sin \theta _1\sin \theta _2A(p,\theta _1)A^*(p,\theta _2)}
{\sqrt {[\cos \theta -\cos (\theta _1+\theta _2)] 
[\cos (\theta _1 -\theta _2) -\cos \theta ]}}+F(p,\theta ).
\label{unit}
\end{equation}
\end{widetext}
The region of integration in (\ref{unit}) is given by the conditions
\begin{equation}
\vert \theta _1 -\theta _2\vert\leq \theta,       \quad
\theta \leq \theta _1 +\theta _2 \leq 2\pi -\theta.
\label{integr}
\end{equation}
The integral term represents the two-particle intermediate states of the 
incoming particles. \correction{ The function $F(p,\theta )$ represents the shadowing 
contribution of the inelastic processes to the elastic scattering amplitude. 
Following Van Hove \cite{hove} it is called the overlap function.
It determines the shape of the diffraction peak and is completely 
non-perturbative. Only some phenomenological models pretend to describe it} 
(see also 
\cite{ads} where its shape is obtained using the unitarity relation in 
combination with experimental data).

Now, let us consider the integral term $I_2$ outside the diffraction peak.
Because of the sharp fall-off of the amplitude (\ref{ampl}) with angle, the 
principal contribution to the integral arises from a narrow region  near 
the line $\theta _1 +\theta _2 \approx \theta $. Therefore one of the amplitudes
should be inserted at small angles within the cone while another one is kept 
at angles outside it. At the beginning, let us  neglect the real parts of 
the amplitude both in the diffraction region and at large angles.  We 
insert Eq. (\ref{ampl}) for one of the amplitudes in $I_2$ and  integrate
over one of the angles.  Then the linear integral equation is obtained:
\begin{eqnarray}
{\rm Im}A(p,\theta )=\frac {p\sigma _t}{4\pi \sqrt {2\pi B}}\int _{-\infty }
^{+\infty }d\theta _1 e^{-Bp^2(\theta -\theta _1)^2/2} {\rm Im}A(p,\theta _1)
 \nonumber \\
+F(p,\theta ).\qquad
\label{linear}
\end{eqnarray}
It can be solved analytically (for more details see \cite{anddre, anddre1})
with the assumption that the role of the overlap function $F(p,\theta )$ is 
negligible outside the diffraction cone\footnote{The results of the paper 
\cite{ads} (see the Figure in there) as well as our results (see Eq. 
(\ref{overl}) and Fig. 2 below with the subsequent discussion) where the 
elastic rescattering $I_2$ was 
subtracted from experimental data give strong support to this assumption.}.
To account for \correction{ the real part of the amplitude}, one replaces
$\sigma _t$ by $\sigma _tf_{\rho }$ where $f_{\rho }=  
1+\rho _d\rho _l$ with average values of ratios of real to 
imaginary parts of the amplitude in and outside the diffraction cone denoted 
as $\rho _d$ and $\rho _l$ correspondingly.
It follows from Eq. (\ref{unit}) that $A_1A_2^*\rightarrow {\rm Im}A_1
{\rm Im}A_2(1+\rho _1\rho _2)$.

Using the Fourier transformation one gets the solution
\begin{eqnarray}
{\rm Im} A(p,\theta )=C_0e^{-\sqrt 
{2B\ln \frac {4\pi B}{\sigma _tf_{\rho }}}p\theta }\nonumber\\
+\sum _{n=1}^{\infty }C_n
e^{-({\rm Re }b_n)p\theta } \cos (\vert {\rm Im }b_n\vert p\theta-\phi _n),
\label{ImA}
\end{eqnarray}

\begin{equation}
b_n\approx \sqrt {2\pi B\vert n\vert}(1+i{\rm sign }n) \qquad n=\pm 1, \pm 2, ...
\end{equation}

This shape has been obtained from contributions due to the pole on the real axis
and a set of the pairs of complex conjugated poles. Correspondingly,
it contains the exponentially decreasing with $\theta $ (or 
$\sqrt {\vert t \vert }$) term (Orear regime!) with imposed on it 
damped oscillations. Let us mention the papers \cite{adya,*adya1} where non-damped
oscillations were predicted in the reggeon exchange model but they are not
observed in experiment.


The elastic scattering differential cross section outside 
the diffraction cone (in the Orear regime region) is
\begin{eqnarray}
\frac {d\sigma }{p_1dt}&= &\left (   e^{-\sqrt 
{2B\vert t\vert \ln \frac {4\pi B}{\sigma _tf_{\rho }}}}
\right.
\nonumber \\
&+&\left.
 p_2e^{-\sqrt {2\pi B\vert t\vert}} \cos (\sqrt {2\pi B\vert t\vert }-\phi)
\right )^2 .    
\label{fit}
\end{eqnarray}
The first (Orear) term is exponentially decreasing with $\theta $ 
(or $\sqrt {\vert t \vert }$) and the second term demonstrates the damped 
($n=1$) oscillations which are in charge of the dip-maximum structure near the
diffraction cone. The omitted terms with larger $n$ in Eq. (\ref{ImA}) are damped stronger because 
they contain $\sqrt n$ in exponents. Let us note that the exponents of the
damped terms are much larger numerically than that of the Orear term
if the experimentally measured values of the diffraction cone slope $B$ and
the total cross section $\sigma _t$ are inserted. Namely $B$ and $\sigma _t$ 
determine mostly the shape of the elastic differential cross section in the 
Orear region between the diffraction peak and the large angle parton scattering. 
The value of $4\pi B/\sigma _t$ is so close to 1 that the first term is very 
sensitive to $\rho _l$.  Thus it becomes possible for the first
time to estimate the ratio $\rho _l$ from fits of experimental data.

Beside the overall normalization constant $p_1$ this formula contains the 
constants $p_2$ and $\phi $ which determine the strength and the phase of the 
oscillation\footnote{The phase was determined in \cite{anddre, anddre1} from the 
iterative solution of the unitarity equation to be equal $\phi \approx \pm \pi 
/4$ (actually with somewhat larger absolute value) but we use it here as a free 
parameter. The first (weaker damped) oscillating term in the exact solution of
the equation (\ref{linear}) has only been taken into 
account in Eq. (\ref{fit}). Let us note the same values of the exponential
damping and the period of the oscillations.}. They can be found from fits of 
experimental data. The constant $p_1$ is determined by the transition point from 
the diffraction cone to the Orear regime. The constants $p_2$ and $\phi $ define 
the depth of the dip and its position.

What concerns the ratios $\rho '$s, one can choose $\rho _d\approx 0.14$ as 
prescribed by the dispersion relations for its value at $t=0$ \cite{drna, blha}
and use $\rho _l$ as another fitted parameter which influences the exponents
in Eq. (\ref{fit}). 

Let us note that all  parameters can depend on energy as well as the
values of the diffraction cone slope $B$ and the total cross section 
$\sigma _t$. Surely, this is unimportant if the fit is done at a fixed energy
as in the present paper.

\correction{ The unitarity condition is not a complete theory. It imposes some 
restrictions on its consequences however. Its solution predicts the dependence 
on $p\theta \approx \sqrt {\vert t\vert }$ but not the dependence on the 
collision energy! Nevertheless, main exponents in (\ref{fit}) depend on energy.
We are able to predict them at different energies if the dependence of the 
diffraction slope $B$ and the total cross section $\sigma _t$ is known from
experiment. In this way different reactions (including $\bar pp$, in particular)
may be analysed.}

Apart from comparison of theoretical predictions with experimental data one can 
get some knowledge about the overlap function $F(p,\theta )$ (see \cite{ads}). 
It is important, in particular, to confirm the assumption about its smallness
outside the diffraction peak.
Then the equation (\ref{unit}) is used as an expression for $F(p,\theta )$:
\begin{widetext}
\begin{equation}
F(p,\theta )= 16p ^2\left (\pi \frac {d\sigma }{dt}/(1+\rho ^2)\right )^{1/2}
-\frac {8p^4(1+\rho _d\rho )}{\pi \sqrt {(1+\rho _d^2)(1+\rho ^2)}}
\int _{-1}^1dz_2\int _{z_1^-}^{z_1^+}dz_1\left [\frac {d\sigma }{dt_1}
\cdot \frac {d\sigma }{dt_2}\right ]^{1/2}K^{-1/2}(z,z_1,z_2),
\label{overl}
\end{equation}
\end{widetext}
where $z_i=\cos \theta _i; \;\;\; K(z,z_1,z_2)=1-z^2-z_1^2-z_2^2+2zz_1z_2$, and
the integration limits are $z_1^{\pm }=zz_2\pm [(1-z^2)(1-z_2^2)]^{1/2}.$

At $\sqrt s=7$ TeV the angles are extremely small so that the kernel becomes 
very singular. $K$ is close to 0 but integrable. The divergence is of the type
$\int dz/\sqrt z$ and can be computed. 
Computing $F$ in the diffraction cone one uses $\rho =\rho _d$.
Outside it  $\rho=\rho _l$.

Let us mention that the inhomogeneous equation (\ref{unit}) has been solved 
\cite{anddre1} by iterations with the overlap function approximated by 
$F/s\sigma _{inel}= \exp [-B_{in}p^2\theta ^2/2]$. Its more precise 
approximation is required to get accurate results but it is important that the
conclusion about the phase $\phi $ remains valid.

Below we show and discuss the obtained results.

\section{A fit of the experimental data}

\begin{figure}
\includegraphics[width=\columnwidth]{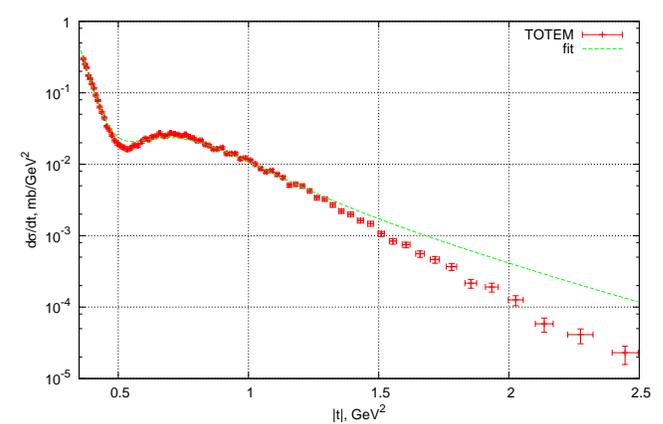}
\caption{Fit of experimental distribution
of elastic $pp$-scattering at $\sqrt s$ = 7 TeV}
\label{totem_fit4}
\end{figure}

Having at our disposal Eq. (\ref{fit}) we try to fit experimental distribution
of elastic $pp$-scattering at $\sqrt s$ = 7 TeV. \correction{ The experimental values of 
$B=20.1$ GeV$^{-2}$ and $\sigma_{t}=98.3$ mb were used in (\ref{fit}).
 We expect that Eq. (\ref{fit}) must be applicable fom the end of the diffraction
cone at $\vert t\vert=0.3$ GeV$^2$ to the beginning of hard parton processes at 
$\vert t\vert>1$ GeV$^2$.}
The result is shown in Fig.~\ref{totem_fit4}.

It is seen that the fit is quite successful in the expected applicability region.
 First of all, we notice
the steeper decrease \correction{ in the region $0.3<\vert t\vert <0.36$ GeV$^2$} 
compared to the slope of the diffraction cone at 
$\vert t\vert < 0.3$ GeV$^2$ as observed in experiment. It is explained here
as the negative contribution of the oscillating term in Eq. (\ref{fit}). That
determines the phase $\phi $. The dip develops at $\vert t\vert $ = 0.53 GeV$^2$
where the cosine in the second term is close to -1. Then this term increases,
becomes  positive and 
leads to the maximum at $\vert t\vert \approx $ 0.7 GeV$^2$. The 
positions of the dip and of the subsequent maximum are uniquely determined by 
the period of oscillations $\Delta t =  2\pi /B$
which is predicted by the unitarity condition and depends only on the well 
measured slope of the diffraction peak $B$. The damping exponent in front of
the $\cos $-term becomes so strong at larger $\vert t\vert $ that the simple 
Orear regime with the first term in Eq. (\ref{fit}) prevails. Let us note that
the exponent in this term is \correction{ rather small because the ratio 
$4\pi B/\sigma _t$ is very close to 1\footnote{This happens
at all energies!}. Therefore it is extremely sensitive to the 
parameter $\rho _l$}. That helps determine this parameter. 

\correction{ Hardly any oscillations will be observed at large $\vert t\vert $. The
exponent in the oscillation term is very large and strongly damps it.} One 
could pretend to observe the next weak oscillation at $\vert t\vert \approx $ 
0.9 - 1.0 GeV$^2$. However it would require very high precision. 
It is interesting to note that the damping increases with energy due to 
increase of the slope $B$. At the same time the shrinkage of the cone leads to 
the shift of the Orear regime (and the dip) to smaller angles at higher energies 
so that the oscillations are stll noticeable there. 

Let us list and discuss the parameters in Eq. (\ref{fit}) which we found by the 
fitting procedure: $p_1=18.71; \; p_2=115.6 ;\;  \phi = -0.845 ;\; 
\rho _l\approx -2$.
\correction{ The large value of $\rho _l$ demonstrates that the dip is well pronounced
in the data.}
Up to now the only possible model-independent estimate of the ratio 
of real to imaginary parts of the elastic scattering amplitude was available from 
the dispersion 
relations at $t=0$. It is for the first time that it is done at large 
$\vert t \vert$ \correction{ in a model-independent way} and shows that this ratio is 
of the order of 1 there. 
\correction{ Surely, there are many models where this ratio is calculated in a wide range of
$t$-values. There is no common consensus about their validity however}.
The parameter $\phi$ is so close to its theoretical 
estimate that it was not even necessary to use it as a free one.

Now we discuss the role of the parameters.

\begin{enumerate} 
\item The parameter $p_1$ is in charge of the overall normalization 
and, consequently, of the smooth transition from the diffraction cone to
the Orear region.

\item The parameter $p_2$ defines the amplitude of the oscillations 
and, consequently, the depth of the dip. In combination with $\phi $ it leads
to the steepened slope at $0.3 < \vert t\vert < 0.36$ GeV$^2$.

\item The phase $\phi $ determines the position of the dip \correction{ and the beginning 
of the steepened slope}.
Actually, it was shown in \cite{anddre1} that it can be obtained from the
iterative solution of the non-linear equation (\ref{unit}). It is almost 
independent of the form of $F(p,\theta )$ so that $\vert \phi \vert \approx
\pi /4$. \correction{ Nevertheless this problem asks for further studies}.

\item The parameter $\rho _l$ in $f_{\rho }$ is in charge of the 
exponential slope at $\vert t\vert $ above the maximum (together with $B$ and 
$\sigma _t$ in the first term of (\ref{fit})).  It is negative and 
rather large (in the absolute value).

\item The relative position of the dip and the maximum (the period of 
oscillations) is determined only by the diffraction cone slope $B$
(the second term in (\ref{fit})).
\end{enumerate}

\section{The overlap function}

\begin{figure}
\includegraphics[width=\columnwidth]{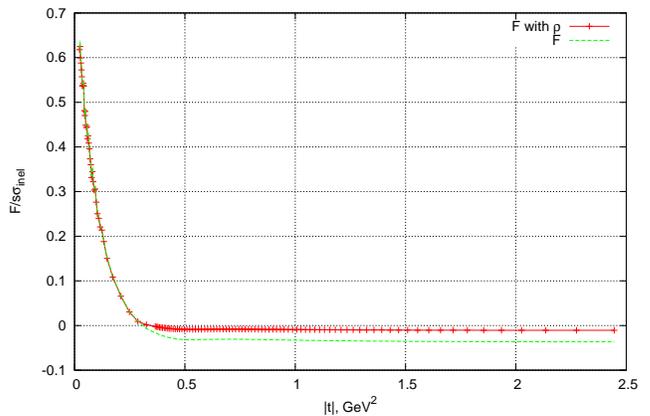}
\caption{The overlap functions calculated with $\rho _d = \rho _l = 0$ and 
with $\rho _d = 0.14; \; \rho _l = -2$ (closest to the abscissa axis).}
 \label{F_rho}
\end{figure}

As follows from experiment, the inelastic cross section is much larger than the 
cross section of elastic scattering at high energies. Therefore the overlap
function is much larger than the integral term in the unitarity relation at
small $t$. To see what is the contribution of inelastic processes to the 
unitarity relation at any values of $t$, it is instructive to calculate the 
overlap function according to Eq. (\ref{overl}). In \cite{ads} that has been 
done at the assumption of ratios of real to imaginary parts $\rho $ equal to 
zero both at small and large $t$. Now with the above estimate of $\rho _l$ we 
can take it into account. Nevertheless, the calculations were done with and 
without account of $\rho $ to compare with previous results and estimate the 
role of $\rho $. The results are shown in Fig.~\ref{F_rho}.

There are several distinctive features observed. First of all, as expected,
the overlap function drops down very fast with increase of the transferred
momentum $\vert t \vert $ and determines the shape of the diffraction cone.
Second, it crosses the abscissa axis at $\vert t\vert \approx 0.3 $ and 
becomes negative. Namely there the Orear regime starts working. If compared to 
low energies \cite{ads}, the overlap function becomes narrower at higher 
energies. Third, it is small and changes very slowly outside the diffraction 
cone similarly to the low energy behavior. \correction{ Intuitively, this smallness
may be understood as a consequence of 
strong destructive interference between amplitudes of inelastic processes with
very different kinematics. In one of these amplitudes the final state must be turned
to the large angle $\theta$ relative to the direction of initial particles.
Thus the overlap of these two processes is small.}
Fourth, the account of $\rho $ does 
not change qualitatively this conclusion in general even though somewhat changes 
the numerical estimates diminishing $\vert F\vert $ further. This follows from
better fit of experimental data with $\rho _l$ different from zero. Fifth, the
negative sign of $F$ imposes a severe problem to theorists because it shows
the important role of the phases of matrix elements of inelastic processes and 
their strong interference when trying to reconstruct elastic scattering from
two inelastic processes turned by $t$ one to another.   

\section{Conclusions}

Thus we conclude:

\begin{itemize}
\item At intermediate angles between the diffraction cone and hard parton
scattering region the unitarity condition predicts the Orear regime with
exponential decrease in angles and imposed on it damped oscillations. Earlier,
this solution was helpful in explaining this regime at lab. energies 8 - 20 GeV.

\item The experimental data on elastic $pp$ differential cross section at
$\sqrt s$=7 TeV in this region are fitted by it with well described 
position of the dip at $\vert t\vert \approx $0.53 GeV$^2$, its 
depth and subsequent damped oscillations with the predicted period about 
0.3 GeV$^2$. The large amplitude of the oscillations and their negative sign 
explain the steepened slope at $0.3 < \vert t \vert <$0.36 GeV$^2$.
The positive sign of the oscillating term at $\vert t \vert \approx$ 0.7 GeV$^2$
leads to the maximum. Strong damping of the oscillations at higher values of 
$\vert t\vert $ results in clear signature of the simple exponential (in 
$\sqrt {\vert t\vert }$) behavior observed first by Orear which extends up to
$\vert t\vert \approx$ 1.5 GeV$^2$.

\item A good fit allows without using any definite model for the first time 
to estimate the ratio of real to imaginary parts of the elastic scattering 
amplitude in this region ($\rho _l\approx -2$) far from forward direction $t$=0.

\item  The overlap function at 7 TeV has been calculated using only the 
experimental differential cross section and the above estimate of the
ratio of real to imaginary parts. As at low energies, it is small 
and negative in the Orear region. That confirms the assumption used 
when solving the unitarity equation and shows that the phases of inelastic 
amplitudes become crucial in any model of inelastic processes.

\end{itemize}

\begin{acknowledgments}
This work was supported by the RFBR grant 09-02-00741 and
by the RAN-CERN program.                                   
\end{acknowledgments}

\bibliography{elast}


\end{document}